\documentclass[10pt]{iopart}
\usepackage{graphicx}
\usepackage{enumerate}
\usepackage{cite}

\usepackage{bm,amssymb}
\begin{document}

\title[Characteristics of the Mott transition]{Characteristics of the Mott transition and electronic states of high-temperature cuprate superconductors from the perspective of the Hubbard model}

\author{Masanori Kohno}

\address{International Center for Materials Nanoarchitectonics (WPI-MANA), 
National Institute for Materials Science, Tsukuba 305-0044, Japan}
\ead{KOHNO.Masanori@nims.go.jp}
\vspace{10pt}
\begin{indented}
\item[]August 2017
\end{indented}

\begin{abstract}
A fundamental issue of the Mott transition is how electrons behaving as single particles carrying spin and charge in a metal change into those 
exhibiting separated spin and charge excitations (low-energy spin excitation and high-energy charge excitation) in a Mott insulator. 
This issue has attracted considerable attention particularly in relation to high-temperature cuprate superconductors, 
which exhibit electronic states near the Mott transition that are difficult to explain in conventional pictures. 
Here, from a new viewpoint of the Mott transition based on analyses of the Hubbard model, 
we review anomalous features observed in high-temperature cuprate superconductors near the Mott transition. 
\end{abstract}

%
%
\submitto{\RPP}
%
\maketitle
%
\ioptwocol

\section{Introduction}
In crystals, electrons usually behave as single particles carrying spin $\hbar/2$ and charge $-e$ under a periodic potential of ions \cite{AshcroftMermin}. 
However, the situation is not necessarily so simple in strongly interacting electron systems, where electronic states are strongly affected by other electrons. 
An obvious example is a Mott insulator \cite{Mott}, which exhibits low-energy spin excitation and high-energy charge excitation: 
the spin and charge excitations are separated. 
Thus, as the electron density increases, the electronic state changes from a metallic state in the low-electron-density regime, where electrons behave as single particles, 
into a Mott insulating state, where spin and charge excitations are separated, because of strong interactions between electrons. 
\par
A natural question is how the electronic state changes. 
As long as we consider that electrons behave as single particles carrying spin and charge, the spin-charge separated excitations in a Mott insulator cannot be well described. 
Conversely, if we consider that the spin and charge degrees of freedom are decoupled, the single-particle behavior in a metal is not well explained. 
This issue of the Mott transition is related to difficulties in dealing with strong electronic correlations (quantum many-body effects), 
and its essence cannot be understood in terms of conventional order parameters for symmetry breaking; 
the Mott transition can occur without magnetic ordering or structural symmetry breaking. 
\par
\begin{figure}
\includegraphics[width=8cm]{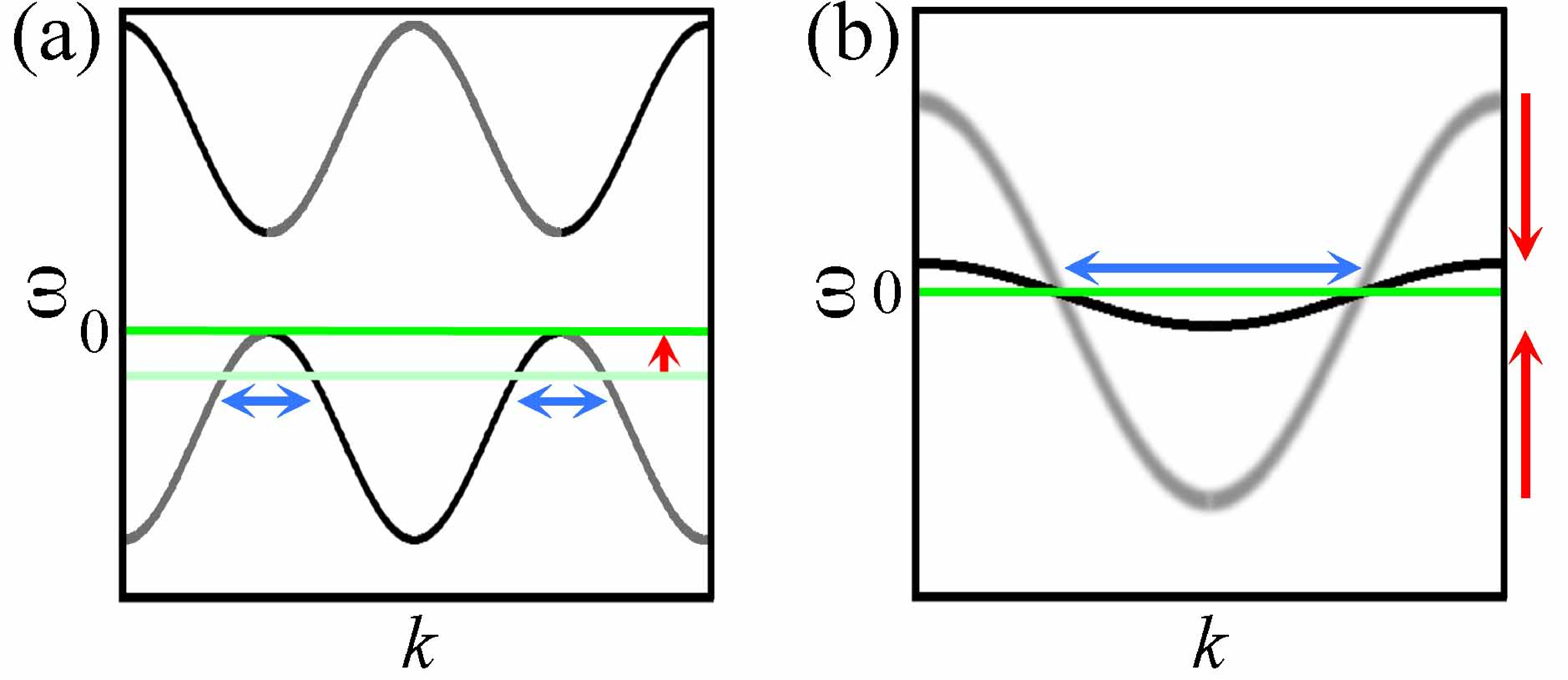}
\caption{Conventional pictures of the Mott transition. 
(a) Schematic band structure (dispersion relation of single-particle excitation) in the rigid-band picture. 
As the chemical potential (green line) increases (red arrow), the volume of hole pockets (light-blue arrows) decreases. 
(b) Schematic band structure in the electron-like quasiparticle picture. 
As the effective mass increases, the bandwidth narrows (red arrows) with the volume inside the Fermi surface (light-blue arrow) remaining large.}
\label{fig:conventional}
\end{figure}

Conventionally, the Mott transition has been considered in the following pictures \cite{ImadaRMP}. 
\begin{itemize}
\renewcommand{\labelitemi}{-}
\item {\it Rigid-band picture.---}The Mott transition is like a metal-insulator transition of a band insulator: 
holes disappear from the band top as the electron density (or the chemical potential) increases toward the Mott transition (figure \ref{fig:conventional}(a)). 
In this case, the carriers are holes. The volume enclosed by the Fermi surface is proportional to the number of holes, which decreases toward the Mott transition. 
\item {\it Electron-like quasiparticle picture.---}Because the Mott transition is caused by strong electronic correlations, the effective mass of electrons (quasiparticles) diverges toward the Mott transition; 
the bandwidth of the quasiparticle narrows and disappears (figure \ref{fig:conventional}(b)). 
In this case, the carriers are the quasiparticles. The volume inside the Fermi surface is proportional to the number of electrons, which remains large even near the Mott transition. 
\end{itemize}
Although these conventional pictures seem to explain some aspects of the Mott transition, some remarkable characteristics cannot be well understood in the conventional pictures. 
In particular, in high-temperature (high-$T_{\rm c}$) cuprate superconductors, which are obtained by doping Mott insulators \cite{BednorzMuller}, 
anomalous electronic states have been observed \cite{ShenRMP,Graf,XrayPRL,UniversalFlatbandBi2212,FlatbandBi2212,LSCO_FS,ArmitageRMP,DagottoRMP,PhillipsRRP}; 
the understanding of the electronic states near the Mott transition is considered key to revealing the mechanism of high-$T_{\rm c}$ superconductivity \cite{AndersonRVB,DagottoRMP,LeeRMP,OgataRev}. 
\par
In this paper, based on recent results for Hubbard-type models \cite{Kohno1DHub,Kohno2DHub,KohnoSpin,Kohno2DHubNN,KohnoDIS,KohnoDIS2,Kohno2DtJ,Kohno2DtJNN,KohnoButsuri}, 
we discuss the nature of the Mott transition 
and review the anomalous electronic states observed in high-$T_{\rm c}$ cuprates from a new viewpoint of the Mott transition. 
\par
The Hubbard model is defined by the following Hamiltonian: 
\begin{equation*}
{\cal H}=-t\sum_{\langle i,j\rangle, \sigma}(c^{\dagger}_{i,\sigma}c_{j,\sigma}+{\rm H.c.})+U\sum_in_{i,\uparrow}n_{i,\downarrow}
-\mu\sum_{i,\sigma}n_{i,\sigma}, 
\end{equation*}
where $\langle i,j\rangle$ indicates that $i$ and $j$ are nearest-neighbor sites. 
Here, $c_{i,\sigma}$ and $n_{i,\sigma}$ denote the annihilation and number operators, respectively, of an electron with spin $\sigma(=\uparrow,\downarrow)$ at site $i$. 
Hereafter, we use units with $\hbar=1$. 
The hole doping concentration is defined as $\delta=1-n$, where $n$ denotes the electron density. 
\par
The ground state of this model at half-filling ($\delta=0$) is a Mott insulator if the Coulomb repulsion $U$ is much larger than the hopping integral $t(>0)$. 
In the Mott insulator, each site is occupied essentially by one electron. Then, if one electron is added, one site is doubly occupied, 
which costs energy of the order of the Coulomb repulsion $U$: charge excitation has a large gap \cite{HubbardI}. 
However, electrons at neighboring sites can exchange spins with a small energy of the order of $J(=4t^2/U)$ through the second-order hopping process \cite{AndersonSE}, 
which implies that there is a low-energy spin excitation. 
Thus, the Mott insulator exhibits separated spin and charge excitations: low-energy spin excitations of $O(J)$ and high-energy charge excitations of $O(U)$. 
\par
To discuss how the electronic state changes when the Mott insulator is doped, we study the spectral function defined as follows: 
\begin{equation*}
A({\bi k},\omega)=\left\{
\begin{array}{lcl}
\frac{1}{2}\sum_{j,\sigma}|\langle j|c^{\dagger}_{{\bi k},\sigma}|{\rm GS}\rangle|^2\delta(\omega-\epsilon_j)&{\rm for}&\omega>0,\\
\frac{1}{2}\sum_{j,\sigma}|\langle j|c_{{\bi k},\sigma}|{\rm GS}\rangle|^2\delta(\omega+\epsilon_j)&{\rm for}&\omega<0,
\end{array}\right.
\end{equation*}
where $c^{\dagger}_{{\bi k},\sigma}$ denotes the creation operator of an electron with momentum ${\bi k}$ and spin $\sigma(=\uparrow,\downarrow)$, 
and $\epsilon_j$ denotes the excitation energy from the ground state $|{\rm GS}\rangle$ to the eigenstate $|j\rangle$. 
The spectral function represents the probability that we can add (remove) an electron with momentum ${\bi k}$ and energy $|\omega|$ for $\omega>0$ ($\omega<0$). 
In the noninteracting case ($U=0$), because $c_{{\bi k},\sigma}^{(\dagger)}|{\rm GS}\rangle$ is an eigenstate, $A({\bi k},\omega)$ shows a single curve, as shown in figure \ref{fig:conventional}(b). 
The spectral function can also be expressed as 
$$A({\bi k},\omega)=-\frac{1}{\pi}{\rm Im}G({\bi k},\omega),$$
by using the retarded single-particle Green function $G({\bi k},\omega)$ \cite{ImadaRMP}. 
The spectral function can be observed in angle-resolved photoemission spectroscopy (ARPES) for $\omega<0$ and in inverse ARPES for $\omega>0$ \cite{ShenRMP}. 

\section{Mott transition in the one-dimensional Hubbard model}
\label{sec:1DHub}
If there is a general characteristic of the Mott transition, it should appear even in the simplest model. 
Hence, in this section, we consider the one-dimensional (1D) Hubbard model, 
for which exact solutions \cite{LiebWu,TakahashiBook,EsslerBook,SchulzSpctra,Kohno1DHub,Benthien} and accurate numerical results are available \cite{Kohno1DHub,Benthien}. 
\par
\begin{figure}
\includegraphics[width=8.5cm]{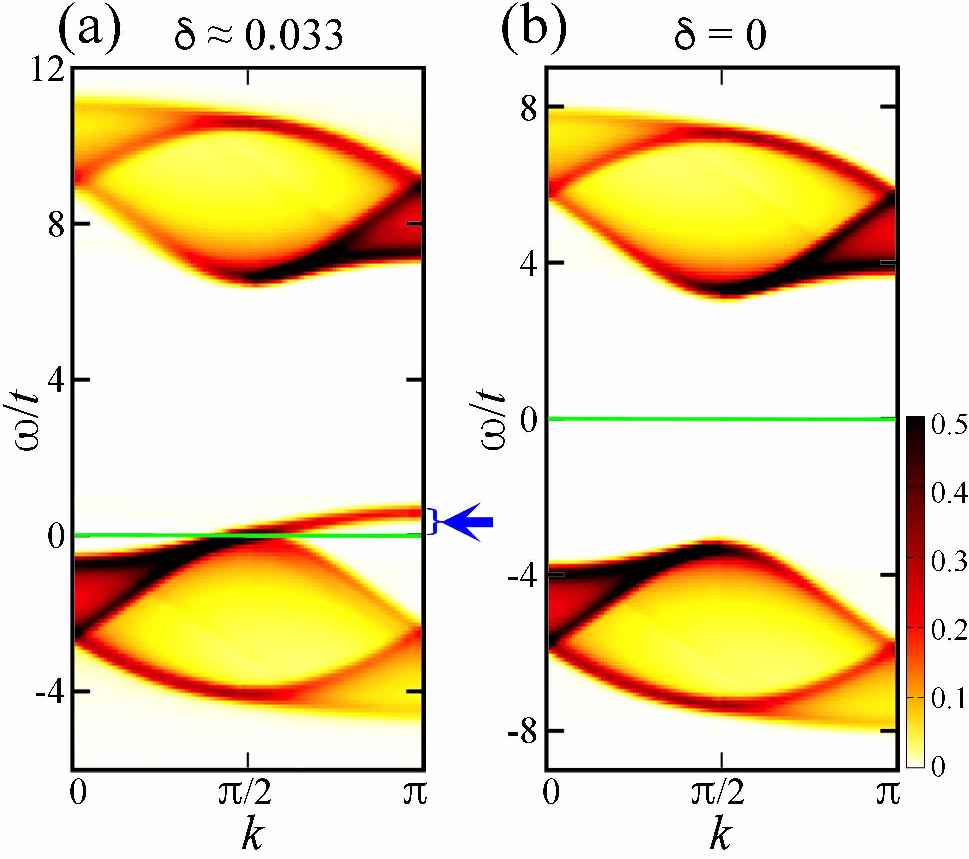}
\caption{Spectral-weight distributions of the 1D Hubbard model for $U/t=10$ obtained using the non-Abelian dynamical density-matrix renormalization group (DDMRG) method 
on a 120-site chain with 120 density-matrix eigenstates \cite{Kohno1DHub,Kohno2DtJ}. 
(a) $A(k,\omega)t$ near the Mott transition ($\delta\approx 0.033$). The blue arrow indicates the doping-induced states. 
(b) $A(k,\omega)t$ of the Mott insulator ($\delta=0$). The green lines indicate the Fermi level ($\omega=0$). 
Gaussian broadening with a standard deviation of $0.1t$ is used.}
\label{fig:1DHub}
\end{figure}

At zero electron density ($n=0$), the spectral function $A(k,\omega)$ exhibits a single mode because there are no other electrons. 
When the system contains more than one electron, spectral weights emerge at high energies, which form the upper Hubbard band (UHB). 
As the electron density increases, the spectral weight of the UHB increases, while that of the lower Hubbard band (LHB) decreases. 
The mode exhibiting a free-electron-like dispersion relation around the Fermi level continuously deforms, decreasing the spectral weight for $\omega>0$ in the LHB \cite{Kohno1DHub}. 
Immediately before the Mott transition, the spectral-weight distribution becomes almost identical to that of the Mott insulator (figure \ref{fig:1DHub}(b)), 
except for the existence of the gapless dispersing mode for $\omega>0$ in the LHB (figure \ref{fig:1DHub}(a)). 
The Mott transition occurs when the chemical potential reaches the top of the LHB at half-filling before reaching the top of the dispersing mode in the small-doping limit. 
Although the dispersing mode loses the spectral weight for $\omega>0$, the dispersion relation retains a wide $\omega$ range even in the small-doping limit \cite{Kohno1DHub}. 
From the insulating side, following the doping of the Mott insulator, 
a gapless dispersing mode emerges in the Mott gap. The spectral weight of this mode increases as the doping concentration increases \cite{Kohno1DHub}. 
\par
This behavior cannot be explained in the conventional pictures. 
In the rigid-band picture, the band structure of the Mott insulator is expected to remain unchanged with or without doping (figure \ref{fig:conventional}(a)): 
the band is simply filled with electrons, and no mode loses the spectral weight. 
In the electron-like quasiparticle picture, the bandwidth near the Mott transition is expected to become extremely narrow as the effective mass diverges (figure \ref{fig:conventional}(b)): 
the dispersion relation of the quasiparticle does not retain a wide $\omega$ range. 
\par
In a mean-field (local spin-density) approximation for an antiferromagnetic long-range order, the effective staggered magnetic field causes the energy gap between the UHB and the LHB, 
and the modes of the single-particle excitation bend back at the boundaries of the magnetic Brillouin zone even in a doped system (figure \ref{fig:conventional}(a)); 
the mode exhibiting a free-electron-like dispersion relation is formed by merging the UHB and the LHB when the antiferromagnetic long-range order disappears. 
However, in the Hubbard model in the large-$U/t$ regime for $0<n<2$, the UHB is separated from the LHB by an energy gap of $O(U)$ 
even without an antiferromagnetic long-range order in the ground state; 
the free-electron-like mode in a doped system is due to the doping-induced states which gain spectral weights as the doping concentration increases (figure \ref{fig:1DHub}(a)). 
\section{Doping-induced states}
\label{sec:DIS}
Although the emergence of states in the Mott gap induced by doping a Mott insulator has been recognized since the early 1990s \cite{Eskes,DagottoDOS,DagottoRMP}, 
its interpretations are controversial. Typically, the doping-induced states have been interpreted as 
\begin{enumerate}[(i)]
\item a part of the UHB, the energy of which is decreased by doping because doping relaxes the binding between double occupancy and vacancy \cite{SakaiPRL}; 
\label{item:DHbinding}
\item states resulting from the hybridization between composite fermions (cofermions) and the quasiparticles \cite{YamajiPRL,YamajiPRB}; 
\label{item:cofermion}
\item states coupled with the charge-$2e$ boson obtained by integrating the high-energy degrees of freedom \cite{PhillipsRMP,PhillipsRRP}; 
\label{item:2eBoson}
\item a spin-polaron shakeoff band which is disconnected from the low-energy quasiparticle band \cite{EderHub, EdertJ}; 
\label{item:spinPolaron}
\item states that lose the spectral weights as the doping concentration decreases and eventually exhibit the momentum-shifted magnetic dispersion relation in the small-doping limit: 
the mode is gapless in the small-doping limit if the spin excitation is gapless in the Mott insulator \cite{Kohno1DHub,Kohno2DHub,KohnoDIS,KohnoDIS2,KohnoSpin,KohnoButsuri,Kohno2DtJ}. 
\label{item:spinExcitedStates}
\end{enumerate}
\par
These interpretations can be classified in terms of their origin and energy gap. 
In interpretations (\ref{item:DHbinding})--(\ref{item:2eBoson}), 
the doping-induced states are related to double occupancy; the mode of the doping-induced states is separated by an energy gap from the low-energy mode. 
In interpretations (\ref{item:spinPolaron}) and (\ref{item:spinExcitedStates}), although the doping-induced states are related to spin excitation, 
the mode of the doping-induced states is separated by an energy gap from the low-energy quasiparticle band in (\ref{item:spinPolaron}) 
but is gapless in (\ref{item:spinExcitedStates}) if the spin excitation of the Mott insulator is gapless. 
\par
Whether the mode of the doping-induced states is disconnected from the low-energy mode or not is important not only for the low-energy electronic properties near the Mott transition 
but also to understand how the free-electron-like electronic state in the large-doping regime changes into the electronic state exhibiting spin-charge separation in the Mott insulator. 
In (\ref{item:DHbinding})--(\ref{item:spinPolaron}), the mode of the doping-induced states is separated from the low-energy mode by an energy gap, 
which is considered responsible for the zero surface of the Green function \cite{SakaiPRL, SakaiPRB, YamajiPRL,YamajiPRB, PhillipsRMP,PhillipsRRP,EderHub}. 
In (\ref{item:spinExcitedStates}), on the other hand, such an energy gap does not exist in the small-doping limit if the spin excitation of the Mott insulator is gapless; 
the free-electron-like mode in the large-doping regime can continuously deform into the mode exhibiting the momentum-shifted magnetic dispersion relation in the small-doping limit. 

\section{Relationship between doping-induced states and spin excitation}
\label{sec:DISspin}
In the 1D Hubbard model, the properties of the doping-induced states have been clarified using exact solutions and accurate numerical results \cite{Kohno1DHub}. 
Here, to explain the essence, we consider the large-$U/t$ regime. 
\par
In the Mott insulator at half-filling, the low-energy properties can be effectively described 
by the Heisenberg model defined by the following Hamiltonian: 
$${\cal H}_{\rm spin}=J\sum_{\langle i,j\rangle}{\bi S}_i\cdot{\bi S}_j,$$ 
where ${\bi S}_i$ denotes the spin operator at site $i$ and $J=4t^2/U$ \cite{AndersonSE}. 
The spin-wave dispersion relation is expressed as 
$$\omega=v_{\rm 1D} |\sin k|,$$
where $v_{\rm 1D}$ denotes the spin-wave velocity of the 1D Heisenberg model ($v_{\rm 1D} =\pi J/2$ \cite{dCP}). 
Although low-energy spin excitation exists, there is no electronic state in the Mott gap (figure \ref{fig:1DHub}(b)). 
\par
\begin{figure}
\includegraphics[width=8.5cm]{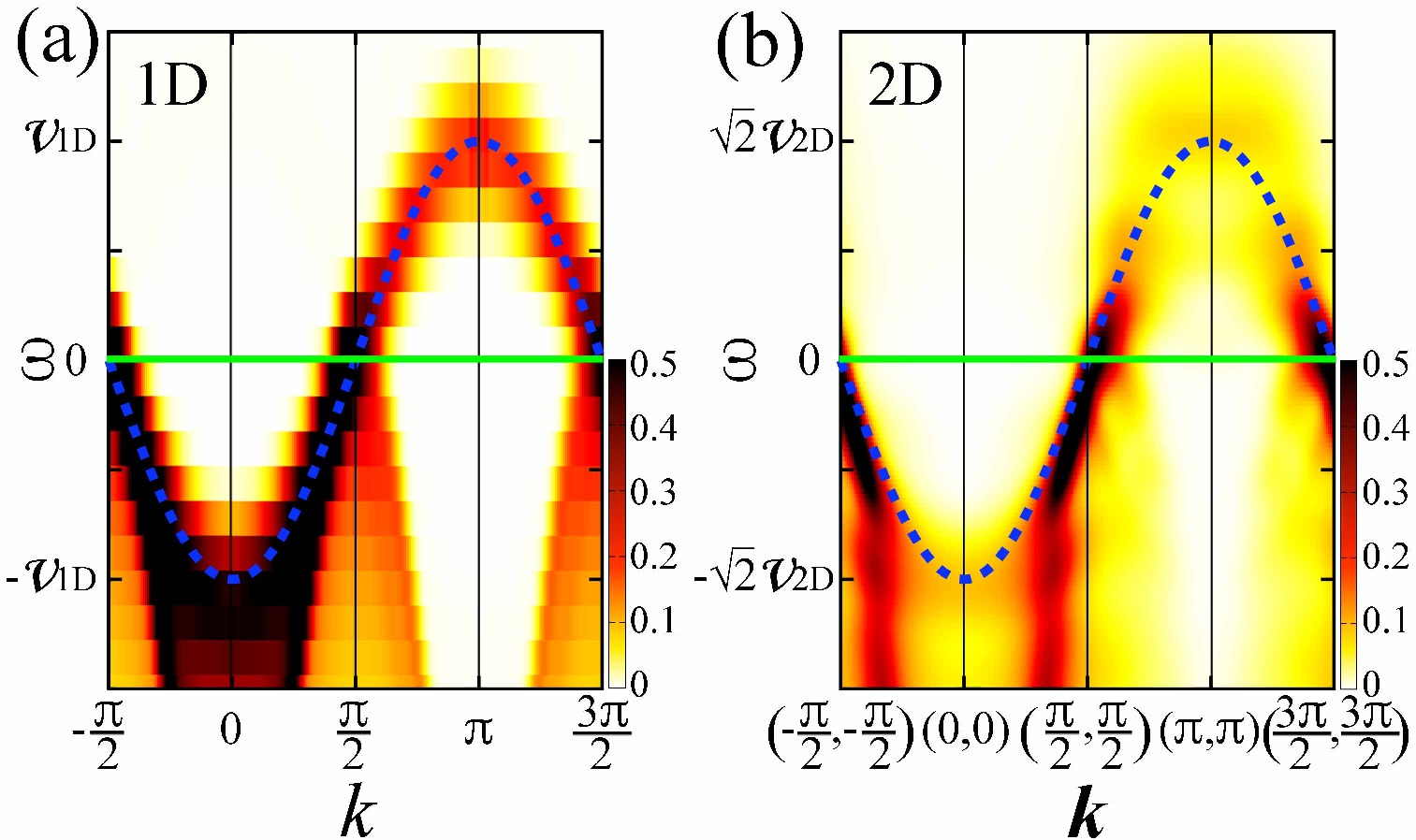}
\caption{Spectral-weight distributions around the Fermi level near the Mott transition in the 1D and 2D Hubbard models for $U/t=10$. 
(a) $A(k,\omega)t$ of the 1D Hubbard model at $\delta\approx 0.033$. Close-up of figure \ref{fig:1DHub}(a) around the Fermi level \cite{Kohno1DHub,KohnoSpin}. 
(b) $A({\bi k},\omega)t$ of the 2D Hubbard model at $\delta= 0.04$ obtained using the cluster perturbation theory for $(4\times 4)$-site clusters \cite{Kohno2DHub,KohnoSpin}. 
The dashed blue curves indicate the spin-wave dispersion relation of the Heisenberg models for $J=4t^2/U$. 
The spin-wave velocities of the 1D and 2D Heisenberg models are denoted by $v_{\rm 1D}$ and $v_{\rm 2D}$, respectively. 
The solid green lines indicate the Fermi level ($\omega=0$). Gaussian broadening with a standard deviation of $0.1t$ is used.}
\label{fig:DIS}
\end{figure}

On decreasing the chemical potential, the Mott transition occurs when the chemical potential reaches the top of the LHB. 
Then, infinitesimal doping induces electronic states in the Mott gap (figure \ref{fig:1DHub}(a)). 
The dispersion relation of the doping-induced states in the small-doping limit can essentially be expressed as 
$$\omega=-v_{\rm 1D} \cos k$$
for $\omega>0$ using the same $v_{\rm 1D}$ as that of the spin-wave dispersion relation (figure \ref{fig:DIS}(a)) \cite{Kohno1DHub}. 
Thus, the doping-induced states exhibit the spin-wave dispersion relation of the Mott insulator with the momentum shifted by the Fermi momentum ($\pi/2$) in the small-doping limit. 
The doping-induced states are directly related to the spin excitation and form a gapless mode reflecting the gapless spin excitation of the Mott insulator. 
As the doping concentration increases, the mode of the doping-induced states continuously deforms into the free-electron-like mode in the large-doping regime. 
\par
From the metallic side, as the electron density increases, the free-electron-like mode in the large-doping regime continuously deforms, gradually losing the spectral weight for $\omega>0$, 
and it eventually exhibits the momentum-shifted spin-wave dispersion relation (figure \ref{fig:1DHub}(a)). 
When the chemical potential is increased beyond the top of the LHB at half-filling, the spectral weight of this mode for $\omega>0$ disappears completely: the Mott insulating state is realized (figure \ref{fig:1DHub}(b)). 
\par
Thus, in the 1D Hubbard model, only interpretation (\ref{item:spinExcitedStates}) 
among the above interpretations ((\ref{item:DHbinding})--(\ref{item:spinExcitedStates}) in section \ref{sec:DIS}) can account for the behavior of the doping-induced states. 
The mode of the doping-induced states is gapless and related to the spin excitation of the Mott insulator as well as to the free-electron-like mode in the large-doping regime. 
\par
The reason why the momentum is shifted by the Fermi momentum can generally be explained as follows \cite{KohnoDIS}. 
At half-filling, we assume that (1) the ground state has spin $S=0$ and momentum ${\bi k}={\bm 0}$, 
(2) the spin excitation with $S=1$ exhibits the dispersion relation $\omega=f({\bi k})$, 
and (3) the top of the LHB is located at ${\bi k}=-{\bi k}_{\rm F}$ (Fermi momentum in the small-doping limit). 
When one electron is removed from the Mott insulator, the ground state should have $S=1/2$ and ${\bi k}={\bi k}_{\rm F}$ because the electron having $S=1/2$ is removed at ${\bi k}=-{\bi k}_{\rm F}$. 
Then, if an electron with momentum ${\bi p}$ is added to the one-hole-doped ground state, 
the electronic state has the same number of electrons as that of the Mott insulator, $S=0$ or $1$ \cite{EdertJ}, and ${\bi k}={\bi p}+{\bi k}_{\rm F}$ \cite{KohnoDIS}. 
The state having $S=0$ and ${\bi k}={\bm 0}$ can have overlap with the ground state of the Mott insulator as in the band insulator case. 
The state having $S=1$ and ${\bi k}={\bi p}+{\bi k}_{\rm F}$ can have overlap with the spin excited state in the Mott insulator at excitation energy $\omega=f({\bi p}+{\bi k}_{\rm F})$; 
$A({\bi k},\omega)$ shows a mode along $\omega=f({\bi k}+{\bi k}_{\rm F})$. 
Thus, the doping-induced states in the small-doping limit exhibit the spin-wave dispersion relation shifted by the Fermi momentum \cite{KohnoDIS}. 
Because this argument relies on neither quasiparticle pictures nor dimensionality, this is a general characteristic of the Mott transition. 
\par
The emergence of the spin excited states in the single-particle spectrum along the momentum-shifted dispersion relation at low energies 
reflects the spin-charge separation of the Mott insulator \cite{KohnoDIS} and contrasts with the band insulator case. 
In a band insulator, the spin-flip excitation costs energy as large as the band gap because spin-charge separation does not occur. 
Thus, by doping a band insulator, no state emerges in the band gap. 
In contrast, in a Mott insulator, there are low-energy spin excited states, which do not appear in the single-particle spectrum, because charge excitation has a much larger energy. 
By doping a Mott insulator, the electron-addition excited states have overlap not only with the ground state of the Mott insulator but also with the low-energy spin excited states, 
which leads to the emergence of spectral weights in the single-particle excitation at low energies along the momentum-shifted magnetic dispersion relation in the Mott gap 
because removing one electron shifts the ground-state momentum by the Fermi momentum. 
\par
In the two-dimensional (2D) Hubbard model, essentially the same characteristic has been obtained in a numerical study \cite{Kohno2DHub}. 
Analogously to the above argument for the 1D Hubbard model, we consider the large-$U/t$ regime, assuming that $U/t$ is not so large as to stabilize ferromagnetic states \cite{Nagaoka}. 
In the Mott insulator of the 2D Hubbard model, the low-energy properties are effectively described by the 2D Heisenberg model, which exhibits 
the spin-wave dispersion relation \cite{AndersonSW} expressed as 
$$\omega=\sqrt{2}v_{\rm 2D} |\sin k|$$ 
for ${\bi k}=(k,k)$ in the $(0,0)$--$(\pi,\pi)$ direction, where $v_{\rm 2D}$ denotes the spin-wave velocity of the 2D Heisenberg model ($v_{\rm 2D}=1.18(2)\sqrt{2}J$ \cite{Singh}). 
In the small-doping regime, the doping-induced states for ${\bi k}=(k,k)$ in the $(0,0)$--$(\pi,\pi)$ direction exhibit the dispersion relation 
$$\omega\approx -\sqrt{2}v_{\rm 2D} \cos k$$
for $\omega>0$ (figure \ref{fig:DIS}(b)), which can essentially be regarded as the magnetic dispersion relation of the Mott insulator 
shifted by the Fermi momentum ${\bi k}_{\rm F}\approx (\pm\pi/2,\pm\pi/2)$ \cite{Kohno2DHub}. 
This is also consistent with the above quantum-number argument.
\par
The reason why almost no spectral weights of the doping-induced states appear in the momentum region inside the Fermi surface in figures \ref{fig:1DHub}(a) and \ref{fig:DIS} is 
that the LHB in this momentum region is almost completely filled with electrons in the ground state of the doped system \cite{KohnoDIS}: there is almost no room for the added electron inside the Fermi surface in the low-energy regime. 
As the doping concentration increases, the number of vacancies, which the added electron can enter with energies much smaller than $U$ 
in the momentum region outside the Fermi surface, increases. 
Thus, increasing the spectral weights in the low-energy regime outside the Fermi surface, the mode of the doping-induced states deforms into the free-electron-like mode carrying considerable spectral weights in the large-doping regime. 
\par
As mentioned in section \ref{sec:DIS}, there have also been numerical or approximate results 
from which other interpretations have been derived for 2D Hubbard-type models 
((\ref{item:DHbinding})--(\ref{item:spinPolaron}) in section \ref{sec:DIS}) \cite{SakaiPRB,SakaiPRL,YamajiPRB,YamajiPRL,PhillipsRMP,PhillipsRRP,EderHub,EdertJ}. 
\par
In the following, we consider how the characteristics of the electronic state near the Mott transition are related to the anomalous features observed 
in high-$T_{\rm c}$ cuprates \cite{ShenRMP,Graf,XrayPRL,UniversalFlatbandBi2212,FlatbandBi2212,LSCO_FS,ArmitageRMP,DagottoRMP,PhillipsRRP}. 

\section{Anomalous spectral features observed in high-temperature cuprate superconductors}
\label{sec:highTc}
High-$T_{\rm c}$ cuprates exhibit superconductivity at much higher temperatures 
(the maximum value of the transition temperature  $T_{\rm c}$ observed so far is approximately 133 K at ambient pressure \cite{Hg1212}) 
compared to conventional superconductors (the values of $T_{\rm c}$ of conventional superconductors are typically less than 40 K at ambient pressure). 
The anomalously high value of $T_{\rm c}$ is considered to be due to the anomalous electronic states of high-$T_{\rm c}$ cuprates, which are obtained by doping Mott insulators. 
Thus, the understanding of the electronic states near the Mott transition is considered key to revealing the mechanism of high-$T_{\rm c}$ superconductivity \cite{AndersonRVB,DagottoRMP,LeeRMP,OgataRev}. 
\par
The following are typical anomalous features of electronic states observed in high-$T_{\rm c}$ cuprates: 
\begin{enumerate}[(a)]
\item {\it Doping-induced states.---}Spectral weights are induced in the Mott gap by doping Mott insulators \cite{XrayPRL,PhillipsRRP}. 
\label{item:DIS}
\item {\it Spinon-like branch and holon-like branch.---}The dispersion relation below the Fermi level in hole-doped systems 
shows two branches similar to the spinon and holon modes of 1D systems \cite{Graf}. 
\label{item:spinonHolon}
\item {\it Giant kink and waterfall.---}The dispersion relation below the Fermi level in hole-doped systems exhibits a kink. 
The dispersion relation becomes steeper, and the spectral weights are significantly reduced immediately below the kink \cite{Graf}.
\label{item:kink}
\item {\it Flat band.---}The dominant mode around ${\bi k}\approx(\pm\pi,0)$ and $(0, \pm\pi)$ exhibits a flat dispersion relation in a wide momentum region 
below the Fermi level near the Mott transition in hole-doped systems \cite{UniversalFlatbandBi2212,FlatbandBi2212}. 
\label{item:flatBand}
\item {\it Pseudogap.---}The spectral weight is reduced near the Fermi level \cite{ShenRMP,DagottoRMP,LeeRMP}. 
\label{item:pseudogap}
\item {\it Fermi arc.---}The spectral weights of some parts of the Fermi surface are significantly reduced near the Mott transition. 
In hole-doped systems, spectral weights near the Fermi level are primarily located only around $(\pm\pi/2,\pm\pi/2)$ \cite{FlatbandBi2212,LSCO_FS}. 
In contrast, those in electron-doped systems are primarily located only around $(\pm\pi,0)$ and $(0, \pm\pi)$ \cite{ArmitageRMP}. 
\label{item:FermiArc}
\end{enumerate}

\section{Single-particle spectral properties of the two-dimensional Hubbard model}
\label{sec:2DHub}
We interpret the above anomalous features from the viewpoint of the proximity of the Mott transition based on recent results for 2D Hubbard models \cite{Kohno2DHub,Kohno2DHubNN}. 
Without loss of generality, we primarily consider the properties for $0\le k_y\le k_x\le \pi$ on a square lattice. 
In this section, we consider the hole-doped case of the 2D Hubbard model; 
the spectral function in the electron-doped case can be obtained by reversing the spectral function in the hole-doped case with respect to $\omega=0$ and ${\bi k}=(\pi/2,\pi/2)$ in the 2D Hubbard model. 
\subsection{Properties along $(0,0)$--$(\pi,\pi)$}
\label{sec:nodal}
\begin{figure}
\includegraphics[width=8.5cm]{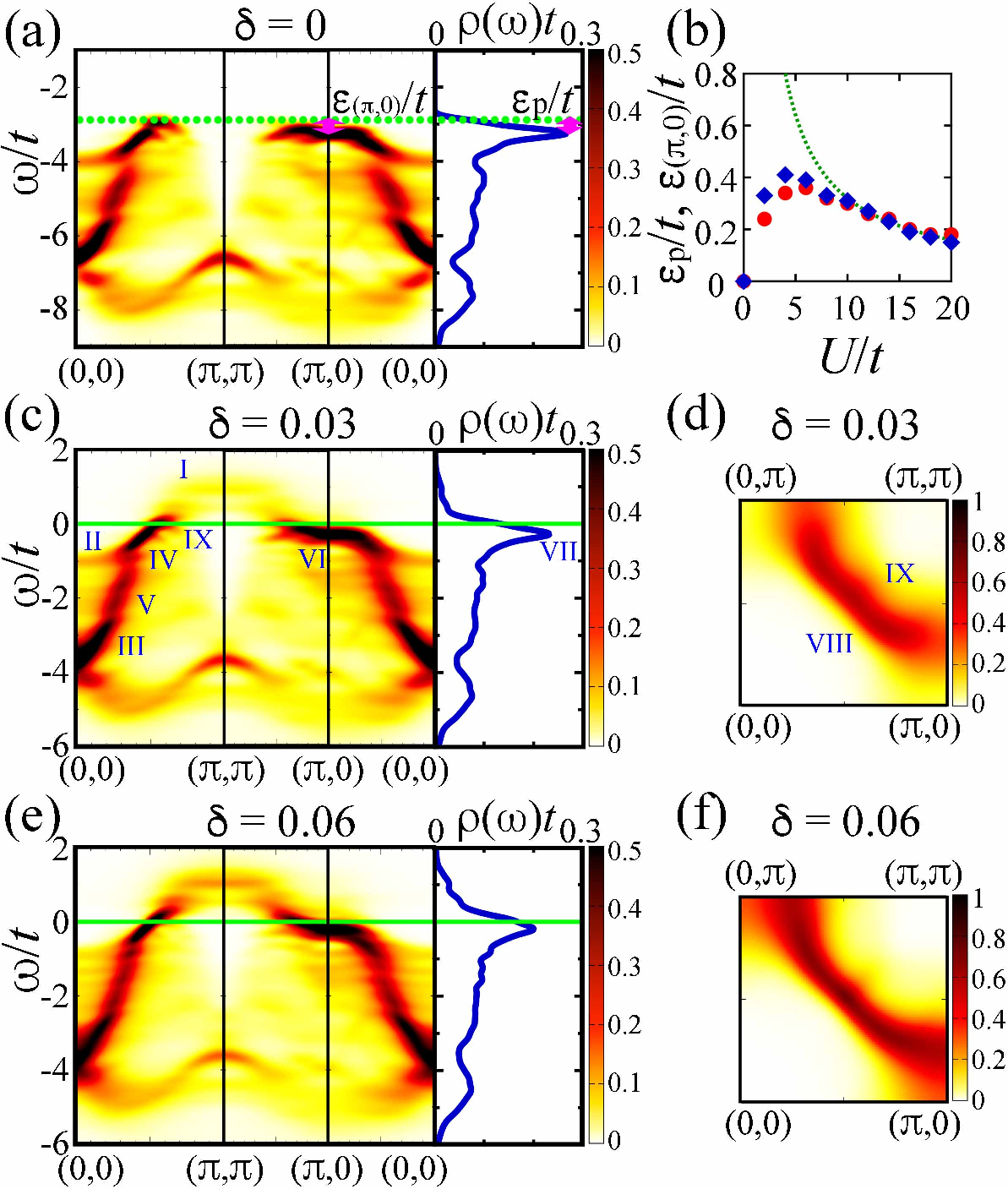}
\caption{Spectral-weight distributions in the LHB of the 2D Hubbard model for $U/t=10$ obtained using the cluster perturbation theory for $(4\times 4)$-site clusters \cite{Kohno2DHub}. 
(a) $A({\bi k},\omega)t$ at $\delta= 0$. The dotted green line indicates the top of the LHB. 
The pseudogap defined by the flat mode at $(\pi,0)$ and that by the main peak of $\rho(\omega)$ are denoted by $\varepsilon_{(\pi,0)}$ and $\varepsilon_{\rm p}$, respectively. 
(b) $\varepsilon_{\rm p}/t$ (blue diamonds) and $\varepsilon_{(\pi,0)}/t$(red circles), taken from Ref. \cite{Kohno2DHub}. 
The dotted green curve indicates a fit in the large-$U/t$ regime, assuming $\varepsilon_{\rm p}, \varepsilon_{(\pi,0)} \propto J(=4t^2/U)$. 
(c) $A({\bi k},\omega)t$ at $\delta= 0.03$. (d) $A({\bi k},\omega\approx 0)t$ at $\delta= 0.03$. 
In (c) and (d), I--IX denote spectral features. I: doping-induced states, II: spinon-like branch, III: holon-like branch, IV: giant kink, V: waterfall, VI: flat mode, VII: pseudogap, VIII: Fermi arc, and IX: hole-pocket-like behavior. 
(e) $A({\bi k},\omega)t$ at $\delta= 0.06$. (f) $A({\bi k},\omega\approx 0)t$ at $\delta= 0.06$. 
In (a), (c), and (e), the rightmost panels show $\rho(\omega)t(\equiv \int d{\bi k}A({\bi k},\omega)t/(2\pi)^2)$. The solid green lines in (c) and (e) indicate the Fermi level ($\omega=0$). 
Gaussian broadening with a standard deviation of $0.1t$ is used in (a) and (c)--(f).}
\label{fig:2dHub}
\end{figure}

We first consider the spectral properties along $(0,0)$--$(\pi,\pi)$. 
As discussed in section \ref{sec:DISspin}, the low-energy spin excited states of the Mott insulator generally emerge in the electron-addition spectrum ($\omega>0$) on doping a Mott insulator, 
and the spectral weight increases as the doping concentration increases. 
Thus, the doping-induced spectral weights observed in high-$T_{\rm c}$ cuprates ((\ref{item:DIS}) in section \ref{sec:highTc}) \cite{XrayPRL} 
can be interpreted as the doping-induced states characteristic of the Mott transition (figure \ref{fig:DIS}(b); I in figure \ref{fig:2dHub}(c)). 
\begin{figure}
\includegraphics[width=8.5cm]{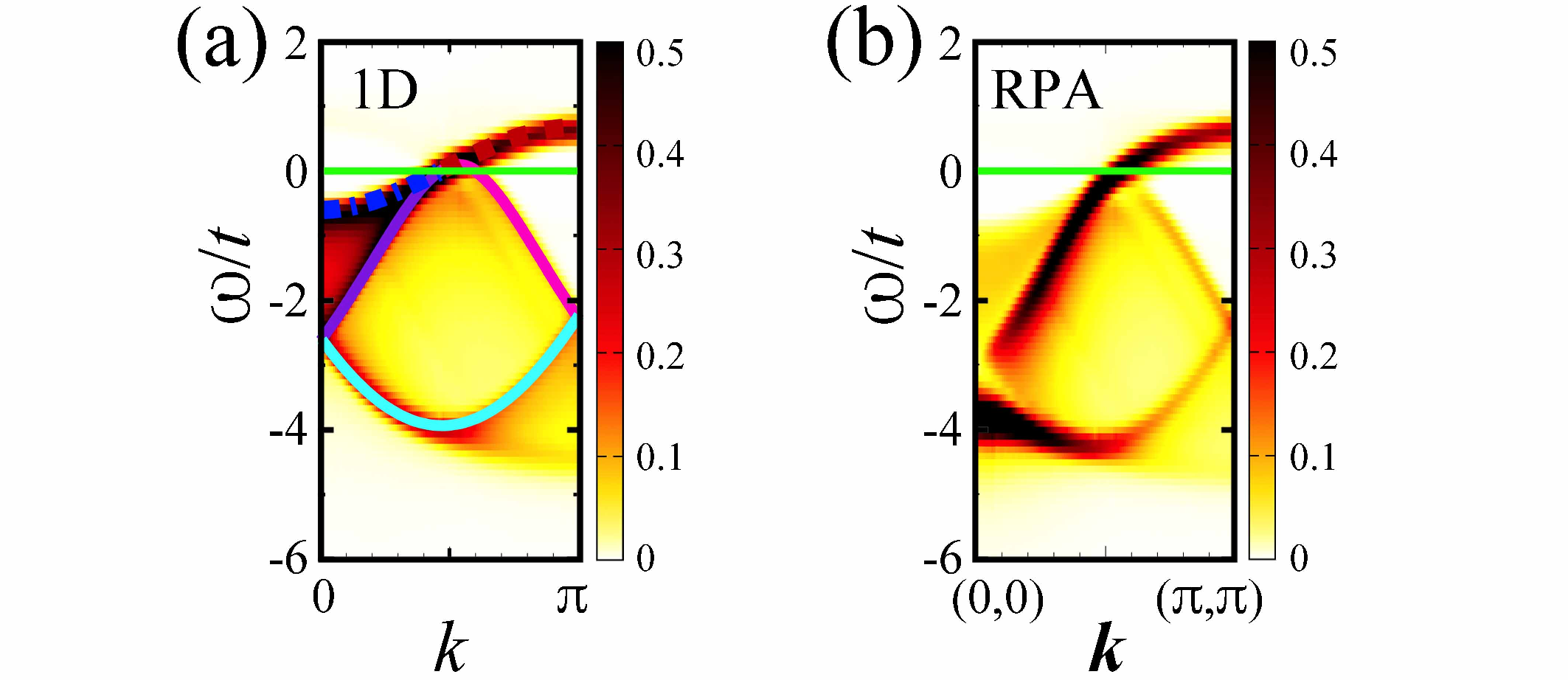}
\caption{Dominant modes of the 1D Hubbard model and their shift by interchain hopping. (a) $A(k,\omega)t$ in the LHB of the 1D Hubbard model for $U/t=10$ at $\delta\approx 0.067$ 
obtained using the non-Abelian DDMRG method on a 120-site chain with 120 density-matrix eigenstates \cite{Kohno1DHub,Kohno2DtJ}. 
The dashed red curve for $\omega>0$ indicates the upper edge of the spinon-antiholon continuum, which corresponds to the doping-induced states. 
The dash-dotted blue curve indicates the spinon mode. The solid curves indicate the holon ($\omega<0$) and antiholon ($\omega>0$) modes. 
(b) $A({\bi k},\omega)t$ along $(0,0)$--$(\pi,\pi)$ obtained by the RPA for interchain hopping ($t_{\perp}=t$) using the data in (a) \cite{Kohno2DHub}. 
The solid green lines at $\omega=0$ indicate the Fermi level. 
Gaussian broadening with a standard deviation of $0.1t$ is used.}
\label{fig:1dRPA}
\end{figure}
\par
We turn our attention to the spectral properties for the electron-removal excitation ($\omega<0$). 
In 1D systems, the electron-removal excitation primarily exhibits two modes: 
the spinon and holon modes (figure \ref{fig:1dRPA}(a)) \cite{Kohno1DHub,Benthien,SchulzSpctra}, which primarily reflect the spin and charge degrees of freedom, respectively. 
The bandwidths of the spinon and holon modes are essentially proportional to the spin exchange coupling $J$  and the hopping integral $t$, respectively; 
the former is about the value of the spin-wave velocity ($\approx\pi J/2$), and the latter, defined by $|\omega|$ at $k=0$, is approximately $2 t$ near the Mott transition. 
\par
This feature is modified by introducing interchain hopping. 
If the interchain hopping integral $t_{\perp} (>0)$ is small, its effects can be investigated using the perturbation theory. 
By assuming that the chains are aligned in the $x$-direction and that they are coupled in the $y$-direction, 
the inverse of the single-particle Green function can be approximated up to the first order in $t_{\perp}$ \cite{RPAArrigoni,RPAWen} as 
$$G^{-1}({\bi k},\omega)\approx G_{\rm 1D}^{-1}(k_x,\omega)-t_{\perp}(k_y),$$
where $G_{\rm 1D}(k_x,\omega)$ denotes the Green function of a 1D chain and $t_{\perp}(k_y)=-2t_{\perp}\cos k_y$. 
In this approximation, which is called the random-phase approximation (RPA), 
the spectral weights are shifted to lower (higher) values of $\omega$ in the momentum region of $t_{\perp}(k_y)<0$ ($t_{\perp}(k_y)>0$) by interchain hopping; 
the shift becomes large for modes carrying large spectral weights \cite{KohnoQ2DHeis,KohnoQ2DHeisH}. 
\par
From this viewpoint, we can infer how the spectral weights shift as interchain hopping is introduced near the Mott transition (figure \ref{fig:1dRPA}(b)) \cite{Kohno2DHub}. 
Around $(0,0)$, spectral weights are expected to shift significantly to lower values of $\omega$ 
because there are considerable spectral weights around $k_x\approx 0$ in the LHB of the 1D Hubbard model (figure \ref{fig:1dRPA}(a)) and because $t_{\perp}(k_y\approx 0)\approx -2t_{\perp}<0$. 
On the other hand, around $(\pi,\pi)$, we expect that spectral weights shift to higher values of $\omega$ because $t_{\perp}(k_y\approx \pi)\approx 2t_{\perp}>0$, 
but the shift would be small because the spectral weights are relatively small around $k_x\approx \pi$ in the LHB of the 1D Hubbard model  (figure \ref{fig:1dRPA}(a)). 
Around $(\pi/2,\pi/2)$, we expect that spectral weights remain almost unaffected by the interchain hopping because $t_{\perp}(k_y\approx \pi/2)\approx 0$. 
\par
The overall behavior of the spectral-weight distribution of the 2D Hubbard model can be interpreted basically from this viewpoint \cite{Kohno2DHub}. 
There are two branches around $(0,0)$ for $\omega<0$: 
the higher-$\omega$ and lower-$\omega$ branches can be interpreted as those originating primarily from the spinon and holon modes, respectively (II and III in figure \ref{fig:2dHub}(c)). 
They are shifted to lower values of $\omega$ from the corresponding 1D modes because of the interchain hopping ($t_{\perp}(k_y\approx 0)<0$). 
At the bifurcation point of these two branches, the dispersion relation exhibits a kink (IV in figure \ref{fig:2dHub}(c)) because their bandwidths are different: 
the bandwidth of the higher-$\omega$ branch ($\approx \sqrt{2}v_{\rm 2D}$; figure \ref{fig:DIS}(b)) is primarily determined by the spin-wave velocity, 
whereas the bandwidth of the lower-$\omega$ branch ($\approx 4t$) is primarily determined by the hopping integral. 
Although the RPA is too simple to explain the reduction in spectral weight immediately below the kink, 
the numerical results for the 2D Hubbard model obtained using the cluster perturbation theory, in which fluctuations within clusters are treated exactly \cite{CPTPRL,CPTPRB}, 
show behavior similar to the waterfall behavior observed in high-$T_{\rm c}$ cuprates (V in figure \ref{fig:2dHub}(c)) \cite{Kohno2DHub}. 
Thus, the spinon-like branch, holon-like branch, giant kink, and waterfall behavior observed in high-$T_{\rm c}$ cuprates ((\ref{item:spinonHolon}) and (\ref{item:kink}) in section \ref{sec:highTc}) 
can be accounted for as properties of a 2D system near the Mott transition. 
\par
The branches responsible for the giant kink and waterfall behavior have also been interpreted as the low-$|\omega|$ renormalized quasiparticle band 
and the high-$|\omega|$ damped spin fluctuation continuum \cite{ScalapinoKink}, string-like excitations \cite{ManusakisKink}, or an incoherent band \cite{ZemljicFiniteT}. 
Moreover, there have been interpretations based on composite-field pictures \cite{PhillipsRRP,AvellaCOM}. 
\par
From the viewpoint of the weakly coupled chains, we expect that the properties around $(\pi/2,\pi/2)$ remain almost unaffected by the interchain hopping. 
This suggests that hole-pocket-like behavior persists around $(\pi/2,\pi/2)$: 
the mode originating from the holon and antiholon modes is expected to bend back around $(\pi/2,\pi/2)$ (IX  in figures \ref{fig:2dHub}(c) and \ref{fig:2dHub}(d)). 
However, the spectral weights around the Fermi level outside the Fermi surface would become very small as the doping concentration increases. 

\subsection{Properties around $(\pi,0)$}
\label{sec:antinodal}
We next consider the spectral properties around $(\pi,0)$. 
In this momentum region, the flat mode carrying large spectral weights is dominant \cite{DagottoRMP}, 
and it is located slightly below the Fermi level near the Mott transition (VI in figure \ref{fig:2dHub}(c)) \cite{Kohno2DHub,Bulut,PreussPG}. 
In the noninteracting case ($U=0$), this mode causes the Van Hove singularity at the Fermi level at half-filling. 
When the interaction is turned on, the Mott gap opens, 
and the mode around $(\pi,0)$ becomes flat in a wider momentum region ((\ref{item:flatBand}) in section \ref{sec:highTc}). 
Because the top of the LHB at half-filling is located near $(\pi/2,\pi/2)$ (figure \ref{fig:2dHub}(a)), 
this mode is located slightly below the Femi level near the Mott transition. 
Because this flat mode carries large spectral weights, it substantially contributes to the main peak of the single-particle density of states. 
The existence of the flat mode slightly below the Fermi level near the Mott transition implies that the density of states is relatively reduced near the Fermi level (VII in figure \ref{fig:2dHub}(c)) \cite{Kohno2DHub}. 
This behavior is referred to as a pseudogap ((\ref{item:pseudogap}) in section \ref{sec:highTc}). 
Although there are several kinds of pseudogaps observed in high-$T_{\rm c}$ cuprates \cite{ShenRMP,DagottoRMP,LeeRMP}, 
here, we consider the pseudogap defined by the main peak of the single-particle density of states to discuss the overall spectral feature. 
The pseudogap energy is defined by $|\omega|$ of the main peak of the density of states, 
which is essentially the same as $|\omega|$ of the flat mode around $(\pi,0)$ in the small-doping limit (figure \ref{fig:2dHub}(b)). 
The numerical results indicate that the pseudogap energy is almost proportion to $J(=4t^2/U)$ in the large-$U/t$ regime (figure \ref{fig:2dHub}(b)), 
which implies that it is related to the antiferromagnetic fluctuation \cite{Kohno2DHub,PreussPG}. 
As the electron density (or the chemical potential) decreases, the pseudogap decreases (figure \ref{fig:2dHub}(e)) 
and closes at a $\delta$ value where the main peak of the density of states or the flat mode around $(\pi,0)$ crosses $\omega=0$. 
\par
Because the flat mode is located around $(\pi,0)$ below the Fermi level near the Mott transition, there is no mode crossing the Fermi level along $(0,0)$--$(\pi,0)$ (figures \ref{fig:2dHub}(c) and \ref{fig:2dHub}(d)). 
In addition, the spectral weights along $(\pi,0)$--$(\pi,\pi)$ near the Fermi level should be significantly reduced because the spectral weights disappear toward the Mott transition. 
Thus, the spectral weights near the Fermi level primarily remain only around $(\pi/2,\pi/2)$ near the Mott transition. 
This feature is referred to as Fermi arc behavior (VIII in figure \ref{fig:2dHub}(d); (\ref{item:FermiArc}) in section \ref{sec:highTc}). 
Accordingly, the Fermi arc behavior can be explained 
as a consequence of the shift of the flat mode below the Fermi level and the reduction in spectral weight of the mode crossing the Fermi level near the Mott transition \cite{Kohno2DHub}. 
As the doping concentration increases, the portion of the Fermi surface having a large spectral weight becomes longer (figure \ref{fig:2dHub}(f)), 
which leads to the free-electron-like Fermi surface in the large-doping regime. 

\section{Effects of next-nearest-neighbor hopping}
\label{sec:NNNhopping}
\begin{figure}
\includegraphics[width=8cm]{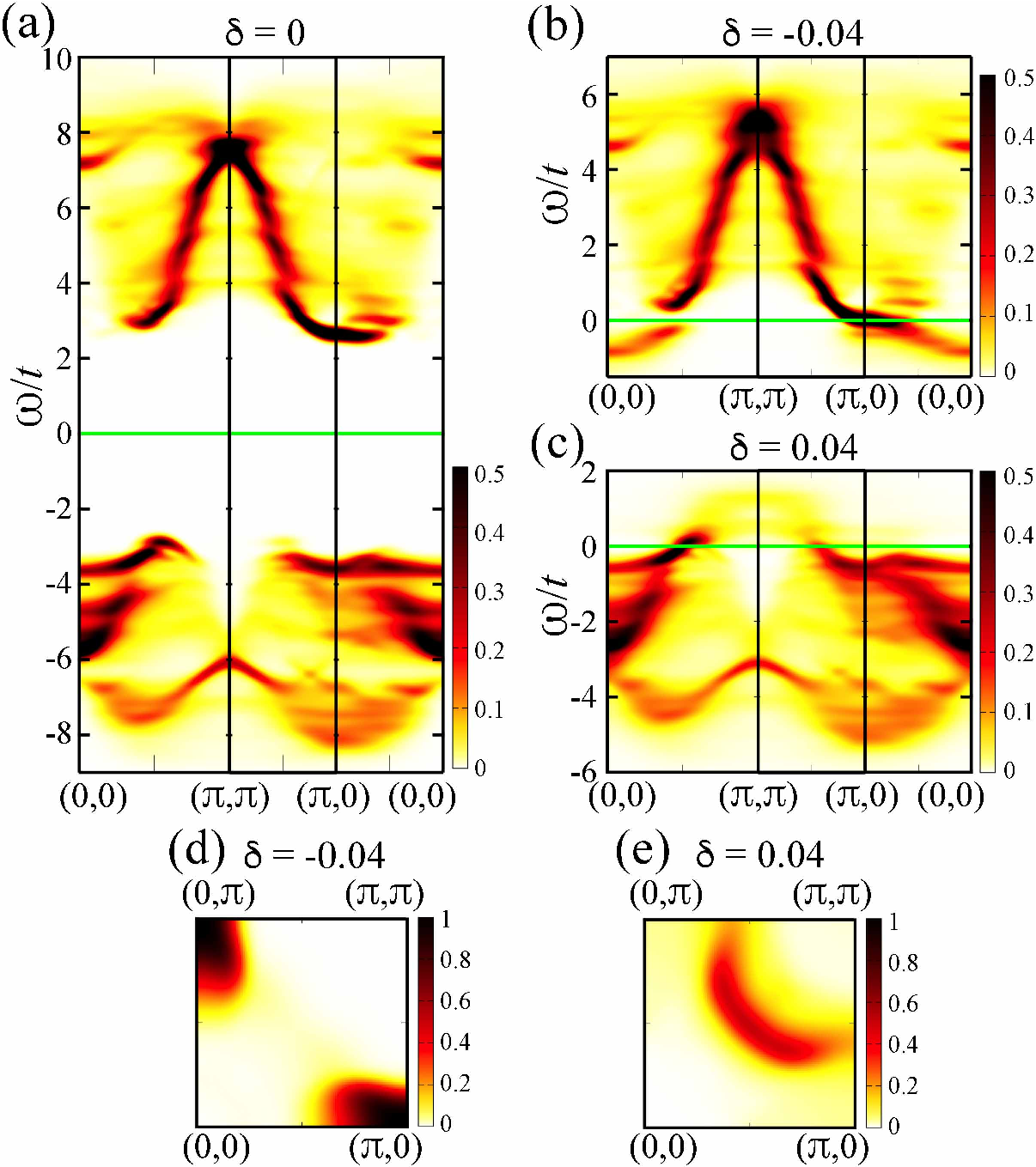}
\caption{Spectral-weight distributions of the 2D Hubbard model with next-nearest-neighbor hopping for $U/t=10$ and $t^{\prime}/t=0.3$ 
obtained using the cluster perturbation theory for $(4\times 4)$-site clusters \cite{Kohno2DHubNN}. 
(a) $A({\bi k},\omega)t$ at $\delta=0$. (b) $A({\bi k},\omega)t$ in the UHB at $\delta= -0.04$. (c) $A({\bi k},\omega)t$ in the LHB at $\delta= 0.04$. 
(d) $A({\bi k},\omega\approx 0)t$ at $\delta= -0.04$. (e) $A({\bi k},\omega\approx 0)t$ at $\delta= 0.04$. 
The green lines in (a)--(c) indicate the Fermi level ($\omega=0$). Gaussian broadening with a standard deviation of $0.1t$ is used.}
\label{fig:2dHubNN}
\end{figure}

To account for the asymmetry of electronic states between hole-doped and electron-doped high-$T_{\rm c}$ cuprates \cite{ShenRMP,ArmitageRMP}, we consider the effects of next-nearest-neighbor hopping. 
The next-nearest-neighbor hopping term is defined by 
$${\cal H}_{\rm nnn}=t^{\prime}\sum_{\langle\langle i,j\rangle\rangle,\sigma}(c^{\dagger}_{i,\sigma}c_{j,\sigma}+{\rm H.c.}),$$
where ${\langle\langle i,j\rangle\rangle}$ indicates that $i$ and $j$ are next-nearest-neighbor sites. 
The numerical results for the 2D Hubbard model with next-nearest-neighbor hopping are shown in figure \ref{fig:2dHubNN}. 
At half-filling, the spectral-weight distribution becomes asymmetric between the LHB and the UHB because of the $t^{\prime}$ term (figure \ref{fig:2dHubNN}(a)). 
In particular, the bottom of the UHB is located near $(\pi,0)$ for moderate values of $t^{\prime}(\approx 0.3t)$, whereas the top of the LHB remains near $(\pi/2,\pi/2)$, 
which causes significant difference in the Fermi surface in the small-doping regime: 
electrons primarily enter the momentum region around $(\pi,0)$ in the electron-doped case, 
whereas holes primarily enter the momentum region around $(\pi/2,\pi/2)$ in the hole-doped case (figures \ref{fig:2dHubNN}(b)--\ref{fig:2dHubNN}(e)). 
This feature has been observed in numerical calculations for Hubbard-type models 
with next-nearest-neighbor hopping \cite{Kohno2DHubNN,CPTPRB,CPTtttU,Tohyama2DtJ,tJTohyamaMaekawa,ZemljicFS,KyungCDMFT,CivelliCDMFT,DahnkenVCPT,eledopeDCA,SakaiPRB,Kohno2DtJNN} 
and in experiments for high-$T_{\rm c}$ cuprates \cite{FlatbandBi2212,LSCO_FS,ArmitageRMP}.
\par
The effects of the $t^{\prime}$ term on the overall spectral features can be interpreted using the RPA-type approximation \cite{Kohno2DHubNN}. 
Analogously to the RPA for interchain hopping (section \ref{sec:nodal}), the inverse of the single-particle Green function is approximated as 
$$G^{-1}({\bi k},\omega)\approx G_{\rm 2D}^{-1}({\bi k},\omega)-t^{\prime}({\bi k}),$$
where $G_{\rm 2D}({\bi k},\omega)$ denotes the Green function of the 2D Hubbard model without next-nearest-neighbor hopping 
and $t^{\prime}({\bi k})=4t^{\prime}\cos k_x\cos k_y$. 
In this approximation \cite{Kohno2DHubNN}, the spectral weights are shifted to lower (higher) values of $\omega$ in the momentum region of $t^{\prime}({\bi k})<0$ ($t^{\prime}({\bi k})>0$) 
by next-nearest-neighbor hopping, and the shift becomes large for modes carrying large spectral weights, as in the RPA for interchain hopping (section \ref{sec:nodal}). 
\par
According to this argument, the spectral weights around $(0,0)$ and $(\pi,\pi)$ are shifted to higher values of $\omega$, 
whereas those around $(\pi,0)$ are shifted to lower values of $\omega$ particularly for modes carrying large spectral weights. 
Thus, the Fermi-arc behavior and pseudogap behavior caused by the flat mode around $(\pi,0)$ are enhanced in the hole-doped case (figures \ref{fig:2dHubNN}(c) and \ref{fig:2dHubNN}(e)). 
In contrast, in the electron-doped case for moderate values of $t^{\prime}(\approx 0.3t)$, because the flat mode around $(\pi,0)$ is shifted by next-nearest-neighbor hopping to lower values of $\omega$ 
than the bottom of the UHB of the 2D Hubbard model at half-filling around $(\pi/2,\pi/2)$, considerable spectral weights are located around $(\pi,0)$ near the Fermi level 
near the Mott transition (figures \ref{fig:2dHubNN}(b) and \ref{fig:2dHubNN}(d)) ((\ref{item:FermiArc}) in section \ref{sec:highTc}) \cite{Kohno2DHubNN}. 
In this interpretation, it is not necessary to assume antiferromagnetic long-range order in hole-doped and electron-doped systems, and the characteristic electronic states near the Mott transition, 
such as the doping-induced states, can collectively be explained as shifted states from those of the 2D Hubbard model (section \ref{sec:2DHub}; figure \ref{fig:2dHub}) 
by next-nearest-neighbor hopping \cite{Kohno2DHubNN}. 
\par
As for the electronic states near the Mott transition in electron-doped systems, there have also been interpretations 
based on antiferromagnetic long-range order \cite{KuskoMF,KusunoseWeakU}. 

\section{Discussion and summary}
In this paper, a new picture of the Mott transition regarding how electrons behaving as single particles in a metal change into those exhibiting the spin-charge separation of the Mott insulator 
was presented \cite{Kohno1DHub,Kohno2DHub,KohnoSpin,KohnoDIS,KohnoDIS2,Kohno2DtJ,KohnoButsuri}. 
The Mott transition is characterized as the freezing of the charge degrees of freedom while the spin degrees of freedom remain active: 
the mode of the electron-addition excitation in the LHB loses the spectral weight while the dispersion relation deforms into the momentum-shifted magnetic dispersion relation of the Mott insulator. 
This is also reasonable in view of the fact that the Mott transition is caused by the Coulomb repulsion, i.e., repulsive interaction between charges. 
From the insulating side, this characteristic can be described as the emergence of spin excitation in the electron-addition spectrum with the dispersion relation shifted by the Fermi momentum 
because the charge character is added by doping. 
Because the dispersion relation of the doping-induced states is shifted from that of the spin excitation by the Fermi momentum, 
the mode of the doping-induced states can be gapless if the spin excitation is gapless. 
As the doping concentration increases, this mode deforms into the free-electron-like mode in the large-doping regime, increasing the spectral weight. 
This characteristic of the Mott transition reflects the spin-charge separation (existence of low-energy spin excitation despite a large charge gap) in a Mott insulator. 
The rigid-band picture and the electron-like quasiparticle picture cannot explain this characteristic 
because these conventional pictures do not properly describe the spin-charge separation of the Mott insulator. 
\par
This characteristic of the Mott transition is relevant to spectral features observed in high-$T_{\rm c}$ cuprates; 
the reduction in spectral weight is related to the doping-induced states and to the Fermi arc. 
The strong low-energy spin fluctuation near the Mott transition is related to the bifurcation of the mode of single-particle excitation into the spinon-like branch and holon-like branch, 
which causes the giant kink and waterfall behavior. 
The pseudogap and the flat mode around $(\pi,0)$ are also related to the antiferromagnetic fluctuation, because their energies are almost proportional to $J(=4t^2/U)$ in the large-$U/t$ regime 
in the 2D Hubbard model. 
Thus, various spectral features observed in high-$T_{\rm c}$ cuprates, which seemed to be anomalous in conventional pictures, 
are explained as properties of a 2D system near the Mott transition \cite{Kohno2DHub}. 
\par
The asymmetric spectral-weight distribution between hole-doped and electron-doped systems can be explained by considering 
how the spectral-weight distribution of the 2D Hubbard model is shifted by next-nearest-neighbor hopping \cite{Kohno2DHubNN}. 
The spectral-weight distribution along $(0,0)$--$(\pi,\pi)$ in the 2D Hubbard model can further be explained by considering how interchain hopping 
modifies the spectral-weight distribution of the 1D Hubbard model \cite{Kohno2DHub,Kohno1DHub}. 
Thus, the overall spectral features of 2D systems near the Mott transition can be interpreted by tracing the origins to those of 1D systems. 
\par
Because essentially the same spectral features have been obtained in the $t$-$J$ models \cite{Kohno2DtJ,KohnoDIS,KohnoDIS2,Kohno2DtJNN}, 
the spectral features discussed in this paper are not specific to the Hubbard models. 
The existence of double occupancy is not essential for these features, because double occupancy does not exist in the $t$-$J$ models. 
\par
To clarify properties in the very small-$|\omega|$ regime near the Mott transition, further studies are necessary; 
the possibility of $d$-wave superconductivity or antiferromagnetic long-range order in the ground state of 2D Hubbard-type models is not ruled out. 
\par
In this paper, overall spectral features near the filling-controlled continuous Mott transition were discussed to clarify the nature of the Mott transition. 
In real materials, Mott transitions can be of the first order because of effects that were not considered in this paper. 
Furthermore, the ground state near the Mott transition can be a $d$-wave superconducting state or an antiferromagnetically ordered state. 
Nevertheless, signatures of the overall spectral features near the Mott transition discussed in this paper would generally appear primarily at energies of $O(J)$, as observed in high-$T_{\rm c}$ cuprates. 
In particular, the emergence of electronic states exhibiting the momentum-shifted magnetic dispersion relation following doping, 
which reflects the spin-charge separation of a Mott insulator, is a general and fundamental characteristic of the Mott transition 
that is elusive in conventional single-particle pictures. 
\ack
This work was supported by JSPS KAKENHI Grant Numbers 23540428 and 26400372, and the World Premier International Research Center Initiative (WPI), MEXT, Japan.

\end{document}